\documentstyle[12pt]{article}

\def\labelmark{}

\def\void{}
\newenvironment{formula}[1]{\def\labelname{#1}
\ifx\void\labelname\def\junk{\begin{displaymath}}
\else\def\junk{\begin{equation}\label{\labelname}}\fi\junk}%
{\ifx\void\labelname\def\junk{\end{displaymath}}
\else\def\junk{\end{equation}}\fi\junk\labelmark\def\labelname{}}
{\ifx\void\labelname\def\junk{\end{array}\end{displaymath}}
\else\def\junk{\end{array}\right.\end{equation}}
\fi\junk\labelmark\def\labelname{}\def\junk{}
}
\newenvironment{formulae}[1]{\def\labelname{#1}
\ifx\void\labelname\def\junk{\begin{displaymath}}
\else\def\junk{\begin{eqnarray}\label{\labelname}}\fi\junk}%
{\ifx\void\labelname\def\junk{\end{displaymath}}
\else\def\junk{\end{eqnarray}}\fi\junk\labelmark\def\labelname{}}

\newcommand{\beq}{\begin{formula}}
\newcommand{\eeq}{\end{formula}}
\newcommand{\beqa}{\begin{formulae}}
\newcommand{\eeqa}{\end{formulae}}
\def\BC{\bb C}
\def\_\BC{\bbi C}

\newcommand{\be}{\beta}

\begin{document}
\begin{titlepage}
\setcounter{page}{1}
\renewcommand{\thefootnote}{\fnsymbol{footnote}}

\begin{flushright}
IC/2000/162\\
hep-th/0010216
\end{flushright}

\vspace{13mm}
\begin{center}
{\Large Thermodynamic Properties of a Quantum 
Group Boson Gas $GL_{p,q}(2)$} 
\vspace{14mm}

{\large Ahmed Jellal$^{}$ 
\footnote{E-mail: jellal@ictp.trieste.it -- jellal@youpy.co.uk }}
\,


\vspace{5mm}
{\em $^{}$ High Energy Physics Section\\
 the Abdus Salam International Centre for Theoretical Physics\\
 Strada Costiera 11, 34100 Trieste, Italy} \\
\vspace{5mm}
\end{center}

\vspace{5mm}

\begin{abstract}
An approach is proposed enabling to effectively describe the behaviour of 
a bosonic system. The approach uses the quantum group $GL_{p,q}(2)$ 
formalism. In effect, considering a bosonic Hamiltonian in terms of the 
$GL_{p,q}(2)$ generators, it is shown that its thermodynamic properties 
are connected to deformation parameters $p$ and $q$. For instance, the
average number of particles and the pressure have been computed. 
If $p$ is fixed to be the same value for $q$, our approach coincides 
perfectly with some results developed recently in this subject. The 
ordinary results, of the present system, can be found when we take the 
limit $p=q=1$.
\end{abstract}

\vspace{8mm}
\vfill
\begin{flushleft}
\end{flushleft}

\end{titlepage}
\newpage

\setcounter{footnote}{0}
\renewcommand{\thefootnote}{\arabic{footnote}}
\renewcommand{\theequation}{\arabic{equation}}
\indent
\indent                                                            %
\section{Introduction}
In the past few years, the application of the quantum groups {\cite{1,2}}
in the study of the physical models has attracted several researchers. 
Indeed, many surprising results have appeared, which involve this 
mathematical background. For example, in quantum Hall effect {\cite{3}}, 
a fascinating relation between deformation parameter of $U_q(sl(2))$ and 
the so-called filling factor is established $[4-10]$. Furthermore, 
it is shown that the deformation gives some information about the 
statistics of anyon systems {\cite{11}} and can solve some 
problems like Azbel-Hofstadter one $[12-14]$.

Another kind of application has appeared recently. 
In effect, the thermodynamic properties of the fermions and bosons are 
investigated in the framework of the quantum groups. For this matter, we
note the works developed in $[15-17]$, where the author has obtained
some more important results, like the generalization of the ordinary physical
quantities of the fermions and bosons. Exactly, he has proved the role that 
quantum group $SU_{q}(2)$ plays in the thermodynamic system at high 
temperature (T), corresponding to the case $z=e^{\be\mu}\ll 1$, where 
$\be={1\over k_BT}$, $k_B$ Boltzman's constant, and $\mu$ the chemical
potential.

This paper is based on the comprehensive work of Ubriaco {\cite{17}},
which we refer to for a list of references concerning the
history of this subject. Thus, we wish to extend the Ubriaco's
theory, developed in {\cite{17}} concerning the bosonic gas, to cover 
some other interesting results. For this, we introduce a 
mathematical background. The latter includes that of quantum group 
$GL_{p,q}(2)$. Otherwise, we are looking for the role that group,
instead of $SU_{q}(2)$, plays in the thermodynamic system at high 
temperature. Effectively, we show that it is possible to obtain the 
thermodynamic properties in terms of two deformation parameters $p$ and 
$q$, when we work with the same case of temperature like in {\cite{17}. 
We mean at high temperature condition, namely $z=e^{\be\mu}\ll 1$. 
Furthermore, we find the results derived in {\cite{17} just by limiting 
to the case $p=q$.

We shall present, in the next section, the quantum group $SU_{q}(N)$-bosons 
background and in particular case where $N=2$. By this we mean a set 
of definitions and notations concerning the generators and $R$-matrix of 
$SU_{q}(N)$. We shall especially insist on the terminology in use in our 
generalization. Section $3$ is devoted to the description of the quantum 
group $GL_{p_{ij},q_{ij}}(N)$-bosons and especially the $GL_{p,q}(2)$ one.
In section $4$, we introduce an explicit physical model. In effect, we 
consider a Hamiltonian, describing two bosonic types, in terms of the 
$GL_{p,q}(2)$ generators.
In that case, we present a new representation of the generators from which 
a new Hamiltonian is obtained. In order to control to what extent or new
Hamiltonian, we compute in section $5$ the thermodynamic properties like 
the average number of particles and the pressure. We conclude and give a brief
comment on the extension of our methods to Bose-Einstein condensation (BEC)
in the final section.

\section{Quantum group $SU_{q}(N)$-bosons}
In this section, we give a short review concerning the quantum group 
$SU_{q}(N)$-bosons, including the generators, $R$-matrix and some elementary
properties. We begin by establishing the commutation relations of $SU_{q}(N)$ 
as follows {\cite{17}}
\begin{equation}
\begin{array}{l}
\Phi_{j}\bar{\Phi}_{i}=\delta_{ij}+qR_{kijl}\bar{\Phi}_{l}\Phi_{k},\\
\Phi_{l}\Phi_{k}=q^{-1}R_{jikl}\Phi_{j}\Phi_{i},\qquad i,j=1,2,...,N,
\end{array}
\end{equation}
which generalize the ordinary ones
\begin{equation}
\phi_{i}\phi_{j}^{+}-\phi_{j}^{+}\phi_{i}=\delta_{ij}.
\end{equation}
$\delta_{ij}=1$ for $i=j$ and $0$ for $i\neq j$. The $R$-matrix is 
{\cite{18}}
\begin{equation}
R_{jikl}=\delta_{jk}\delta_{il}(1+(q-1)\delta_{ij})+
(q-q^{-1})\delta_{ik}\delta_{jl}\theta(j-i),
\end{equation}
and the function $\theta(j-i)$ is given by
\begin{equation}
\theta(j-i)=\left\{
\begin{array}{l}
1\qquad {\mbox{if}}\; j>i,\\
0\qquad {\mbox{otherwise}}.
\end{array}
\right.
\end{equation}
With the help of the following linear transformation 
\begin{equation}
\Phi_i^{'}=\sum_{j=1}^{N}T_{ij}\Phi_i,
\end{equation}
it is shown that eqs.(1) are covariant under the $SU_{q}(N)$ 
transformation. The matrix $T_{ij}$ is the $N$-dimensional representation
of the present group, which verifies the algebraic relations {\cite{19}}
\begin{equation}
\begin{array}{l}
RT_{1}T_{2}=T_{2}T_{1}R,\\
R_{12}R_{13}R_{23}=R_{23}R_{13}R_{12},
\end{array}
\end{equation}
where $T_{1}=T\otimes {\bf 1}$ and $T_{2}={\bf 1}\otimes T\in V\otimes V$ 
and $(R_{23})_{ijk,i^{'}j^{'}k^{'}}=\delta_{ii^{'}}R_{jk,j^{'}k^{'}}
\in V\otimes V\otimes V$. For a given unitary quantum group matrix 
$T=\pmatrix{a&b\cr 
c&d\cr}$ and from $RTT$ equation it is found {\cite{20}}
\begin{equation}
\begin{array}{l}
ab=q^{-1}ba,\;\;\;\;\;\;\; ac=q^{-1}ca,\\
bc=cb,\;\;\;\;\;\;\; dc=qcd,\\
db=qbd,\;\;\;\;\;\;\; da-ad=(q-q^{-1})bc,\\
det_{q}T=ad-q^{-1}bc=1,\\
\end{array}
\end{equation}
where $q\in {\bf R}$ and the adjoint 
matrix of $T$ 
\begin{equation}
{\bar T}=\pmatrix{d&-qb\cr 
-q^{-1}c&a\cr},
\end{equation}
are taken in to consideration.

The commutation relations generating the quantum group 
$SU_q(2)$-bosons are given by
\begin{equation}
\begin{array}{l}
\Phi_{2}\bar{\Phi}_{2}-q^{2}\bar{\Phi}_{2}\Phi_{2}=1,\\
\Phi_{1}\bar{\Phi}_{1}-q^{2}\bar{\Phi}_{1}\Phi_{1}=1+(q^{2}-1)
\bar{\Phi}_{2}\Phi_{2},\\
\Phi_{2}\Phi_{1}=q\Phi_{1}\Phi_{2},\\
\Phi_{2}\bar{\Phi}_{1}=q\bar{\Phi}_{1}\Phi_{2},
\end{array}
\end{equation}
which can be obtained by replacing $N=2$ in eqs.$(1)$.
 
Having given some basic notions related to the quantum group 
$SU_{q}(N)$-bosons. Now we turn to generalize the above definitions and
properties. For this matter, we introduce the multiparametric quantum group 
$GL_{p_{ij},q_{ij}}(N)$ tools in the next section.

\section{Quantum group $GL_{p_{ij},q_{ij}}(N)$-bosons}	
Let us remonte to another quantum group. In effect, we would like to 
establish a set of the background useful in the next, which leads
to obtain a quantum group bosons. For this matter, we start by recalling
$GL_{p_{ij},q_{ij}}(N)$, $i,j=1,...,N$. The latter is defined  
by the commutation relations {\cite{21}}
\begin{equation}
R(T\otimes {\bf 1})({\bf 1}\otimes T)=({\bf 1}\otimes T)
(T\otimes {\bf 1})R,\\
\end{equation}
which can be written otherwise
\begin{equation}
R_{rs}^{ik}T_{v}^{r}T_{w}^{s}=T_{f}^{k}T_{e}^{i}R_{vw}^{ef},
\end{equation}
where $k,r,s,v,w,e,f=1,,...,N$ and the $R$-matrix is {\cite{21}}
\begin{equation}
(R_{p_{ij},q_{ij}})_{kl}^{ij}=\delta_{k}^{i}\delta_{l}^{j}[\delta^{ij}+
{\theta(j-i)\over q_{ij}}+\theta(i-j){1\over p_{ij}}]+
(1-p_{ij}^{-1}q_{ij}^{-1})\delta_{l}^{i}\delta_{k}^{j}\theta(i-j).
\end{equation}
The latter satisfies the Yang-Baxter equation
\begin{equation}
R_{12}R_{13}R_{23}=R_{23}R_{13}R_{12},
\end{equation}
where the $\theta$-function is given by Eq.(4). The inverse of 
$R_{p_{ij},q_{ij}}$ is 
\begin{equation}
(R_{p_{ij},q_{ij}})^{-1}=R_{{1\over p_{ij}},{1\over q_{ij}}}.
\end{equation}

In this step, we would like to introduce a differential calculus for the
reason which will appear in the next. To do this, we would like to remember
a work of Schirrmacher done in {\cite{21}}. In this reference, the author has 
constructed a differential calculus on the $N$-dimensional vector space 
associated to the multiparametric deformation of $GL(N)$, which generalized 
the one-parameter differential calculus on the Manin plane {\cite{22}}.
Indeed, the cordinates $x_{i}$ and the derivatives $\partial_{i}$
verify the following relations 
\begin{equation}
\begin{array}{l}
x_{i}x_{j}=q_{ij}x_{j}x_{i},\\
\partial_{i}\partial_{j}=p_{ij}^{-1}\partial_{j}\partial_{i},\\
\partial_{i}x_{i}=1+p_{ij}q_{ij}x_{i}\partial_{i}+(p_{ij}q_{ij}-1)
\sum_{j=i+1}^{N}x_{j}\partial_{j},\\
\partial_{i}x_{j}=p_{ij}x_{j}\partial_{i},\\
\partial_{j}x_{i}=q_{ij}x_{i}\partial_{j},
\end{array}
\end{equation}
with $i,j=1,..., N$ and $i<j$. We note two remarks. The first one, the above 
differential calculus is covariant under the $GL_{p_{ij},q_{ij}}(N)$
transformation. The second when $N=2$, equations $(15)$ coincide 
perfectly with two-parameter generalization of the Wess and Zumino 
differential calculus on the Manin plane {\cite{23}}.

Now we are looking for a realization of the quantum group bosons. 
To do this let us proceed as follows
\begin{equation}
\begin{array}{l}
x_{i}\longrightarrow\bar{\Phi}_{i},\qquad
\partial_{i}\longrightarrow\Phi_{i}.
\end{array}
\end{equation}
These correspondences lead us to obtain  
\begin{equation}
\begin{array}{l}
\bar{\Phi}_{i}\bar{\Phi}_{j}=q_{ij}\bar{\Phi}_{j}
\bar{\Phi}_{i},\\
\Phi_{i}\Phi_{j}=p_{ij}^{-1}\Phi_{j}\Phi_{i},\\
\Phi_{i}\bar{\Phi}_{i}=1+p_{ij}q_{ij}\bar{\Phi}_{i}\Phi_{i}+
(p_{ij}q_{ij}-1)\sum_{j=i+1}^{N}\bar{\Phi}_{j}\Phi_{j},\\
\Phi_{i}\bar{\Phi}_{j}=p_{ij}\bar{\Phi}_{j}\Phi_{i},\\
\Phi_{j}\bar{\Phi}_{i}=q_{ij}\bar{\Phi}_{i}\Phi_{j},
\end{array}
\end{equation}
which generate the quantum group $GL_{p_{ij},q_{ij}}(N)$-bosons.
The latter admits the following involution (Hermitian conjugation)
\begin{equation}
\begin{array}{l}
\Phi_{i}\longrightarrow\bar{\Phi}_{i},\qquad
q_{ij}\longrightarrow p_{ij}.
\end{array}
\end{equation}
To reproduce eqs.$(1)$, we can fix $p_{ij}=q_{ij}=q$. 

Now we focus ourselves on the interesting case $N=2, q_{ij}=q$ and $p_{ij}$,
which will be useful in the next of our study. Introducing these conditions
in eqs.$(17)$, we find the $GL_{p,q}(2)$ commutation relations
\begin{equation}
\begin{array}{l}
\Phi_{1}\Phi_{2}=p^{-1}\Phi_{2}\Phi_{1},\\
\Phi_{1}\bar{\Phi}_{1}=1+pq\bar{\Phi}_{1}\Phi_{1}+(pq-1)
\bar{\Phi}_{2}\Phi_{2},\\
\Phi_{2}\bar{\Phi}_{2}=1+pq\bar{\Phi}_{2}\Phi_{2},\\
\Phi_{1}\bar{\Phi}_{2}=p\bar{\Phi}_{2}\Phi_{1},\\
\Phi_{2}\bar{\Phi}_{1}=q\bar{\Phi}_{1}\Phi_{2},\\
\bar{\Phi}_{1}\bar{\Phi}_{2}=q\bar{\Phi}_{2}\bar{\Phi}_{1}.
\end{array}
\end{equation}
When $p=q$, these equations coincide perfectly with eqs.$(9)$. The 
algebraic relations analogue to eqs.$(7)$ can be obtained from eq.$(11)$
\begin{equation}
\begin{array}{l}
ab=pba,\;\;\;\;\;\;\; ac=qca,\\
pbc=qcb,\;\;\;\;\;\;\; cd=pdc,\\
bd=qdb,\;\;\;\;\;\;\; ad-da=(p-q^{-1})bc,\\
\end{array}
\end{equation}
corresponding to the quantum group $GL_{p,q}(2)$, where 
$0\leq q<\infty$ and $0\leq p<\infty$.

This concludes this section. In the next section we would like to 
analyse a physical model in the framework of the above tools. For
this, we consider a bosonic Hamiltonian involving the generators of 
the quantum group $GL_{p,q}(2)$. 

\section{$GL_{p,q}(2)$-boson model}
Before going on, we would like to make a remark. Indeed,  
our approach is different from that known in the history of this subject,
for example $[24-27]$. So, our main goal is to generalize the work
done recently in {\cite{17}}. For this matter, let us consider a Hamiltonian
describing two types of bosons with the same energy like
\begin{equation}
H=\sum_{k}\epsilon_{k}({\hat{D}}_{1,k}+{\hat{D}}_{2,k}),
\end{equation}
where the operators ${\hat{D}}_{1,k}$ and ${\hat{D}}_{2,k}$ are given by
\begin{equation}
{\hat{D}}_{1,k}=\bar{\Phi}_{1,k}\Phi_{1,k},\qquad 
{\hat{D}}_{2,k}=\bar{\Phi}_{2,k}\Phi_{2,k},
\end{equation}
and $\epsilon_{k}$ is the spectrum of energy, $k=0,1,2...$ We show 
that the following relations 
are satisfied for a given $k$
\begin{equation}
\begin{array}{l}
{\hat{D}}_{2}\Phi_{1}-\Phi_{1}{\hat{D}}_{2}=0,\\
q{\hat{D}}_{1}\Phi_{1}-p^{-1}\Phi_{1}{\hat{D}}_{1}=0.
\end{array}
\end{equation}
The normalized states of the Hamiltonian eq.$(21)$ can be built as follows 
\begin{equation}
|d_1,d_2>={\bar{\Phi}_{2}^{d_{2}}\bar{\Phi}_{1}^{d_{1}}
\over\sqrt{\{d_{1}\}_{p,q}!\{d_{2}\}_{p,q}!}}|0,0>,
\end{equation}
where $|0,0>$ is the ground state of $H$ and we propose that the symbol 
$\{x\}_{p,q}$ takes the form
\begin{equation}
\{x\}_{p,q}={1-q^{x}p^{x}\over 1-qp},
\end{equation}
and $\{x\}_{(p,q)}!$ is 
\begin{equation}
\{x\}_{p,q}!=\{x\}_{p,q}\{x-1\}_{p,q}...1.
\end{equation}

Our goal in the next section is to compute the thermodynamic
properties of the above Hamiltonian. To do this, let us introduce a
new representation. Indeed, we would like to establish a relation
between the new operators and the old ones, namely the ordinary boson
operators. In order to reply to this request, we propose the following
representation for a given $k$
\begin{equation}
\begin{array}{l}
\Phi_{2}=(\phi_{2}^{+})^{-1}\{D_{2}\}_{p,q},\\
\bar{\Phi}_{2}=\phi_{2}^{+},\\
\Phi_{1}=(\phi_{1}^{+})^{-1}\{D_{1}\}_{p,q}p^{D_{2}},\\
\bar{\Phi}_{1}=\phi_{1}^{+}q^{D_{2}}.
\end{array}
\end{equation}
where
$D_{1,k}=\phi_{1,k}^{+}\phi_{1,k}$, $D_{2,k}=\phi_{2,k}^{+}\phi_{2,k}$ 
and $\phi_i$, ${\phi}_{i}^{+}$ are the ordinary boson operators. 
It easy to show that these equations satisfy the commutation
relations given by eqs.$(19)$. Furthermore, by using the last set of 
equations and eq.$(23)$, we can rewrite the above Hamiltonian as follows
\begin{equation}
H=\sum_{k}\epsilon_{k}\{D_{1,k}+D_{2,k}\}_{p,q}.
\end{equation}
Note that this representation led us to obtain an interacting
Hamiltonian instead of the original one eq.$(21)$.

Having obtained the Hamiltonian in terms of the ordinary boson operators, 
let us now calculate explicity its thermodynamic properties. 

\section{Thermodynamic properties}
In this section, we are looking for the thermodynamic properties of the 
bosonic system descibed by the Hamiltonian eq.$(28)$. To get these properties,
we begin by evaluating the corresponding grand partition function. The 
latter can be written as follows
\begin{equation}
Z= Tr \exp[-\beta\epsilon_{k}(\bar{\Phi}_{1,k}\Phi_{1,k}+
\bar{\Phi}_{2,k}\Phi_{2,k})]\exp[\beta\mu(\bar{\phi}_{1,k}
\phi_{1,k}+\bar{\phi}_{2,k}\phi_{2,k})].
\end{equation}
Using the above tools, we show that the last equation becomes
\begin{equation}
Z=\prod_{k}\sum_{d_{1,k}=0}^{\infty}\sum_{d_{2,k}=0}^{\infty}  
\exp[-\beta\epsilon_{k}\{d_{1,k}+d_{2,k}\}_{p,q}]
\exp[\beta\mu(d_{1,k}+d_{2,k})],
\end{equation}
which can be written otherwise
\begin{equation}
Z=\prod_{k}\sum_{d_{k}=0}^{\infty}(d_{k}+1)  
\exp[-\beta\epsilon_{k}\{d_{k}\}_{p,q}]z^{d_{k}},
\end{equation}
where $z=e^{\beta\mu}$ is called the fugacity. It is clear that a direct 
computation of this equation is very hard. However, there is a limit case 
which leads to obtain the thermodynamic properties of eq.$(28)$. For this, 
let us consider the limit $z\ll 1$, which corresponds to the high temperature 
case. Then, taking into account this approximation, we can write 
eq.$(31)$ as follows 
\begin{equation}
\begin{array}{l}
\ln Z={4\pi V\over h^{3}}\int_{0}^{\infty}p^{2}dp
[2e^{-\beta\epsilon}z+(6e^{-\beta\epsilon\{2\}_{p,q}}
-4e^{-2\beta\epsilon}){z^{2}\over 2}+\\ 
(24e^{-\beta\epsilon\{3\}_{p,q}}-36e^{-\beta\epsilon\{2\}_{p,q}}
+8e^{-3\beta\epsilon}){z^{3}\over 3!}...].
\end{array}
\end{equation}
The manipulation of the integral in the $3$-dimentional momentum space 
leads to 
\begin{equation}
\begin{array}{l}
\ln Z={4\pi V\over h^{3}}[{{\sqrt\pi}\over 2}
({2m\over\beta})^{3\over 2}z
+{\sqrt\pi}({2m\over\beta})^{3\over 2}
\delta(p,q)z^{2}+...],
\end{array}
\end{equation} 
where the function $\delta(p,q)$ is given by
\begin{equation}
\delta(p,q)={1\over 4}[{3\over (1+pq)^{3\over 2}}-{1\over\sqrt 2}].
\end{equation}
We note that when we take the limit $p=q$, we find for the 
grand function partition coincides with the one derived in 
ref.{\cite{17}} concerning the boson gas.

Having obtained eq.$(33)$, we return now to investigate the 
thermodynamic properties related to the bosonic system described by 
the above Hamiltonian. For these, we begin by the average number of 
particles. The latter is given by
\begin{equation}
\langle N\rangle={1\over\beta}({\partial ln Z\over\partial\mu})_{T,V}.
\end{equation} 
Using eq.$(34)$, we find
\begin{equation}
\begin{array}{l}
\langle N\rangle={4\pi V\over h^{3}}[{{\sqrt\pi}\over 2}
({2m\over\beta})^{3\over 2}z+2{\sqrt\pi}({2m\over\beta})^{3\over 2}
\delta(p,q)z^{2}+...].
\end{array}
\end{equation}
By inverting this equation, we prove that the fugacity can be written as
\begin{equation}
z\approx{1\over 2}({h^{2}\over 2m\pi kT})^{3\over 2}{\langle N\rangle
\over V}-\delta(p,q)({h^{2}\over 2m\pi kT})^{3}({\langle N\rangle
\over V})^{2}.
\end{equation} 
In the same way, the pressure is related to the grand function partition by
\begin{equation}
P={1\over\beta}({\partial ln Z\over\partial V})_{\mu,T}.
\end{equation} 
All calculations done, we get 
\begin{equation}
\begin{array}{l}
P={4\pi \over h^{3}\beta}[{{\sqrt\pi}\over 2}({2m\over\beta})^{3\over 2}z
+{\sqrt\pi}({2m\over\beta})^{3\over 2}\delta(p,q)z^{2}+...].
\end{array}
\end{equation}
The above relations led us to obtain an equation of state like
\begin{equation}
PV=kT\langle N\rangle(1-\delta(p,q)[({h^{2}\over 2m\pi kT})^{3\over 2}
{\langle N\rangle\over V}+...].
\end{equation}
Note that this expression depends on deformation parameters
$p$ and $q$.

For a system of particles living in two dimension space, the 
above equation of state can be written as follows  
\begin{equation}
PV=kT\langle N\rangle[1-\eta(p,q){h^{2}\over 2m\pi kT}
{\langle N\rangle\over A}+...],
\end{equation}
where $A$ is the surface in which confined the bosonic system and the 
function $\eta(p,q)$ is 
\begin{equation}
\eta(p,q)={2-pq\over 4(1+pq)}.
\end{equation} 
Note that the last two equations are obtained in the same way
as above for a $3$-dimension space.

This analysis generalizes the bosonic results derived in ref.{\cite{17}}. 
The latter can be found in the limit where $p=q$. Furthermore, for $p=q=1$
we obtain the classical results, i.e. the thermodynamic properties of the 
ordinary bosons.

\section{Conclusion}
In this paper, we have investigated a bosonic system in the framework
of the quantum group $GL_{p,q}(2)$. To do this, we have first introduced
a differential calculus. The latter leads us to obtain the multiparametric
quantum group $GL_{p_{ij},q_{ij}}(N)$ and then the $GL_{p,q}(2)$ one. Second,
we have considered a bosonic Hamiltonian in terms of $GL_{p,q}(2)$
generators. Subsequently, we have calculated the thermodynamic properties
starting from the corresponding grand partition function. Therefore, we have
shown, for example, that the average number of particles and the pressure
depending on deformation parameters $p$ and $q$. The equation of state is also
obtained in this way. This task generalized the recent one {\cite{17}}
concerning the boson gas. The latter results are reproduced when we take the
limit $p=q$ in our analysis. Also the ordinary results can be found for the
case where  $p=q=1$.
 
The low temperature case will be considered in a subsequent work {\cite{28}}.
This will include a study of the Bose-Einstein condensation (BEC) in the
framework of the quantum group $GL_{p,q}(2)$-bosons.

\section*{Acknowledgment}
A. Jellal is grateful to Prof. S. Randjbar-Daemi for the kind hospitality 
at the High Energy Section of the Abdus Salam International Centre for 
Theoretical Physics (AB-ICTP) while this work was being completed.
It is a great pleasure for A. Jellal to thank Prof. G. Thompson for 
reading the present work.

\end{document}